\documentclass[floatfix,aps,amsmath,amssymb,reprint]{revtex4-2}

\usepackage{graphicx}
\usepackage{dcolumn}
\usepackage{bm}
\usepackage{placeins}
\usepackage{float}
\usepackage{appendix}
\usepackage{xcolor}
\begin{document}

\title{Activity gradients in two- and three-dimensional active nematics}%

\author{Liam J. Ruske}
\email{liam.ruske@physics.ox.ac.uk}
\author{Julia M. Yeomans}
\affiliation{Rudolf Peierls Centre For Theoretical Physics, University of Oxford, UK}

\date{\today}

\begin{abstract}
We numerically investigate how spatial variations of extensile or contractile active stress affect bulk active nematic systems in two and three dimensions. In the absence of defects, activity gradients drive flows which re-orient the nematic director field and thus act as an effective anchoring force. At high activity, defects are created and the system transitions into active turbulence, a chaotic flow state characterized by strong vorticity. We find that in two-dimensional (2D) systems active torques robustly align $+1/2$ defects parallel to activity gradients, with defect heads pointing towards contractile regions. In three-dimensional (3D) active nematics disclination lines preferentially lie in the plane perpendicular to activity gradients due to active torques acting on line segments. The average orientation of the defect structures in the plane perpendicular to the line tangent depends on the defect type, where wedge-like $+1/2$ defects align parallel to activity gradients, while twist defects are aligned anti-parallel. Understanding the response of active nematic fluids to activity gradients is an important step towards applying physical theories to biology, where spatial variations of active stress impact morphogenetic processes in developing embryos and affect flows and deformations in growing cell aggregates, such as tumours. 
\end{abstract}

\maketitle


\section{\label{sec:intro}Introduction}

Active systems continuously take energy from their environment and convert it into mechanical stress which drives the system out of equilibrium. 
Active matter has recently received considerable attention in the community because of its potential as a way of interpreting biological mechanics and as examples of non-equilibrium statistical physics \citep{julicher2018hydrodynamic, marchetti2013hydrodynamics, chen2018mechanical}. 

Dense active systems, and in particular the continuum theory of active nematics, have been used successfully to model the dynamics of a variety of biological systems, such as myosin-driven actin-microtubule networks \citep{lee2021myosin}, bacterial biofilms \citep{yaman2019emergence} and confluent cell monolayers \citep{saw2017topological, mueller2019emergence, duclos2017topological, saw2018biological}. A key property of active nematics, which distinguishes them from passive liquid crystals, is active turbulence. This is a chaotic flow state characterised by strong vorticity and topological defects which are continually created and destroyed. Unlike in passive systems, where topological defects of opposite charge tend to annihilate over time due to the relaxation of the large elastic energy associated with them, in active systems the number of topological defects is finite at steady-state, and motile $+1/2$ defects move through the fluid like self-propelled, oriented particles \citep{vromans2016orientational}. Considerable experimental and theoretical work has been devoted to understanding the statistical properties of active turbulence and the motion of the associated topological defects, in both two and three dimensions \citep{duclos2020topological, binysh2020three, vcopar2019topology,ruske2021morphology}. 

A major research theme has been to investigate how to control the chaotic flows and defect motion in active materials. Guiding active flow and defects along selected paths is likely to be a prerequisite to using active matter for
powering micro-machines or targeting the delivery of drugs. It has been proposed that this control can be achieved in two-dimensional (2D) systems by patterns of activity on a substrate. This causes the orientation of $+1/2$ defects to align along activity gradients and hence the defects to move along paths defined by activity patterns \citep{zhang2021spatiotemporal, shankar2019hydrodynamics}. Further work has investigated the alignment of individual topological defects by activity gradients in 2D using theoretical arguments and numerical simulations \citep{tang2021alignment, mozaffari2021defect}. More recently it has also proved possible to design active materials that allow imaging of a three-dimensional (3D) active nematic and, in particular, the associated motile disclination loops and lines \citep{duclos2020topological}. These articles have focused primarily on how activity patterns affect flows and the alignment of individual topological defects, and it is still poorly understood how activity gradients affect systems in the absence of topological defects or how well the concept of 2D topological defects as effective oriented particles translates to disclination line segments in 3D materials. 

In this paper we numerically investigate how active nematics respond to spatial variations in activity in both two and three dimensions. We start by outlining the well-established hydrodynamic theory of active nematics and the numerical methods we use to solve the equations of motion. 
In section~\ref{subsec:act} we show how non-uniform activity creates active forces which, in the absence of defects, set up flows which rotate the nematic director field. This effectively creates an active force which favours either normal or tangential alignment of the nematic director with respect to activity gradients for contractile or extensile active stress, respectively. 
In the following section~\ref{subsec:2d}, we show that activity gradients in 2D active turbulence induce polar order of $+1/2$ defects, which is consistent with previous theoretical predictions considering active torques on isolated defects \citep{tang2021alignment}.
In section~\ref{subsec:3d} we investigate how activity gradients in 3D systems affect nematic alignment and the polarisation of disclination lines. We find that disclination lines, which locally resemble $+1/2$ or twist defects, are polarized in opposite directions, aligning parallel or anti-parallel to activity gradients, respectively. 
The last section of the paper summarizes the key results, highlights similarities between activity gradients and active-passive interfaces and points out possible connections to experiment.

\section{\label{sec:methods} Equations of motion}

We numerically solve the continuum equations of motion of an active nematic fluid \citep{marenduzzo2007hydrodynamics}. The ordering of the nematic constituents in 3D is quantified by the tensor order parameter $Q_{ij} = \frac{3}{2}S  (n_i n_j - \delta_{ij}/3)$, where the scalar $S$ quantifies the magnitude of the order and the director field $\mathbf{n}$ the direction. We work in the uniaxial limit,  ignoring higher-order effects such as biaxial order close to disclination cores \citep{schimming2020anisotropic, long2021geometry}. The evolution of the nematic tensor $\mathbf{Q}$ is coupled to the velocity field $\mathbf{u}$ of the fluid and follows \citep{beris1994thermodynamics}
\begin{equation}
    D_t \mathbf{Q} -\mathbf{\mathcal{W}} = \Gamma \mathbf{H} \: ,
    \label{eq:Q}
\end{equation}
where $D_t$ denotes the material derivative and $\Gamma$ the rotational diffusivity which controls the relaxation of the order parameter towards equilibrium through the molecular field $\mathbf{H} = -\delta \mathcal{F}/\delta \mathbf{Q} + (\mathbf{I}/3) \text{Tr}( \delta \mathcal{F} / \delta \mathbf{Q})$. The free energy $\mathcal{F}=\int f dV$ of the system includes a Landau-de Gennes bulk term $f_{bulk} = \frac{A}{2} \text{Tr}~\mathbf{Q}^2 + \frac{B}{3} \text{Tr}~\mathbf{Q}^3 + \frac{C}{4} (\text{Tr}~\mathbf{Q}^2)^2$, where we chose the coefficients $A,B,C$ such that nematic ordering $S_{eq}>0$ is favoured in equilibrium, and an elastic term $f_{el} = K_{el} \left( \mathbf{\nabla} \mathbf{Q} \right)^2$, which penalizes distortions in the director field $\mathbf{n}$ using the one-elastic-constant approximation. This commonly used approximation has been shown to accurately describe many experimental systems in 2D and 3D \citep{saw2018biological,saw2017topological,norton2018insensitivity,duclos2020topological,zhang2018interplay,schimming2020anisotropic}. It should also be noted that active nematic dynamics are only weakly affected by the bulk energy scale and mainly depend on the magnitude of $K_{el}$. The second term on the left-hand side of eq.~(\ref{eq:Q}) is the co-rotational term $\mathbf{\mathcal{W}}$ which describes the response of a flow-tumbling director field to gradients in the flow field $\mathbf{u}$ \citep{marenduzzo2007hydrodynamics}, 
\begin{equation}
\mathbf{\mathcal{W}}_{ij} = \Omega_{ik} Q_{jk} + Q_{ik} \Omega_{jk} \: ,
\end{equation}
where $\Omega_{ij}=(\partial_j u_i - \partial_i u_j)/2$ is the anti-symmetric part of the velocity gradient tensor. The time evolution of the velocity field obeys the incompressible Navier-Stokes equation
\begin{equation}
    \rho \: D_t \mathbf{u} = \mathbf{\nabla} \cdot \left( \Pi^{visc} + \Pi^{el} + \Pi^{act} \right) \: ,
    \label{eq:u}
\end{equation}
where $\rho$ is the density of the fluid and the stress tensor on the right-hand side comprises the viscous stress $\Pi^{visc}$ of a Newtonian fluid, elastic stress $\Pi^{el}$ caused by the relaxation of $\mathbf{Q}$ (for details see Supplementary Information \citep{si}), and an active stress component $\Pi^{act}$. Activity locally induces dipolar stress, of magnitude $\zeta$, along the axis of the director field, $\Pi^{act}_{ij} = -\zeta Q_{ij}$ which drives the system out of equilibrium \citep{simha2002hydrodynamic}. The parameter $\zeta$ quantifies the type and magnitude of dipolar stress produced by the constituents, where $\zeta>0$ describes extensile systems in which particles push out fluid along their long axis $\mathbf{n}$ and pull in fluid in the plane perpendicular to $\mathbf{n}$. $\zeta<0$ describes contractile activity where the flow direction is reversed. It is apparent from eq.~(\ref{eq:u}) that active forces are induced by gradients in either $\mathbf{Q}$ or $\zeta$. They induce the well-known hydrodynamic bend or splay instability of the director field $\mathbf{n}$ in homogeneous active nematic bulk systems \citep{simha2002hydrodynamic}. We solve the equations of motion using a hybrid Lattice Boltzmann-finite difference method \citep{marenduzzo2007hydrodynamics} with parameter values stated in the Supplementary Information \citep{si}. 

We consider periodic variations of activity along one coordinate axis with alternating extensile ($\zeta>0$) and contractile activity ($\zeta<0$) in two and three-dimensions,
\begin{equation}
    \zeta(\mathbf{x}) = \zeta_{max} \cdot \cos(\mathbf{k} \cdot \mathbf{x}) \: ,
\end{equation}
with wave-vector $\mathbf{k}$ indicating the direction $\mathbf{k}/|\mathbf{k}|$ and wavelength $\lambda = 2\pi/|\mathbf{k}|$ of activity patterns (Fig.~\ref{fig:1} a,f). In the following we will denote the coordinate axes parallel and perpendicular to the plane of constant activity as the tangential and normal direction, respectively. We define a dimensionless activity number $\mathcal{A}=\lambda \sqrt{|\zeta_{max}|/K_{el}}$ as the ratio of the wavelength $\lambda$ to the active nematic length-scale $\ell_\zeta = \sqrt{K_{el}/|\zeta_{max}|}$. For more complex, aperiodic activity patterns $\mathcal{A}$ can be defined as the ratio of a characteristic domain size $L$ to the active length-scale $\ell_\zeta$.

\section{\label{subsec:act} Active forces in the absence of defects}

\begin{figure}
	\centering
	\includegraphics[width=8.7cm]{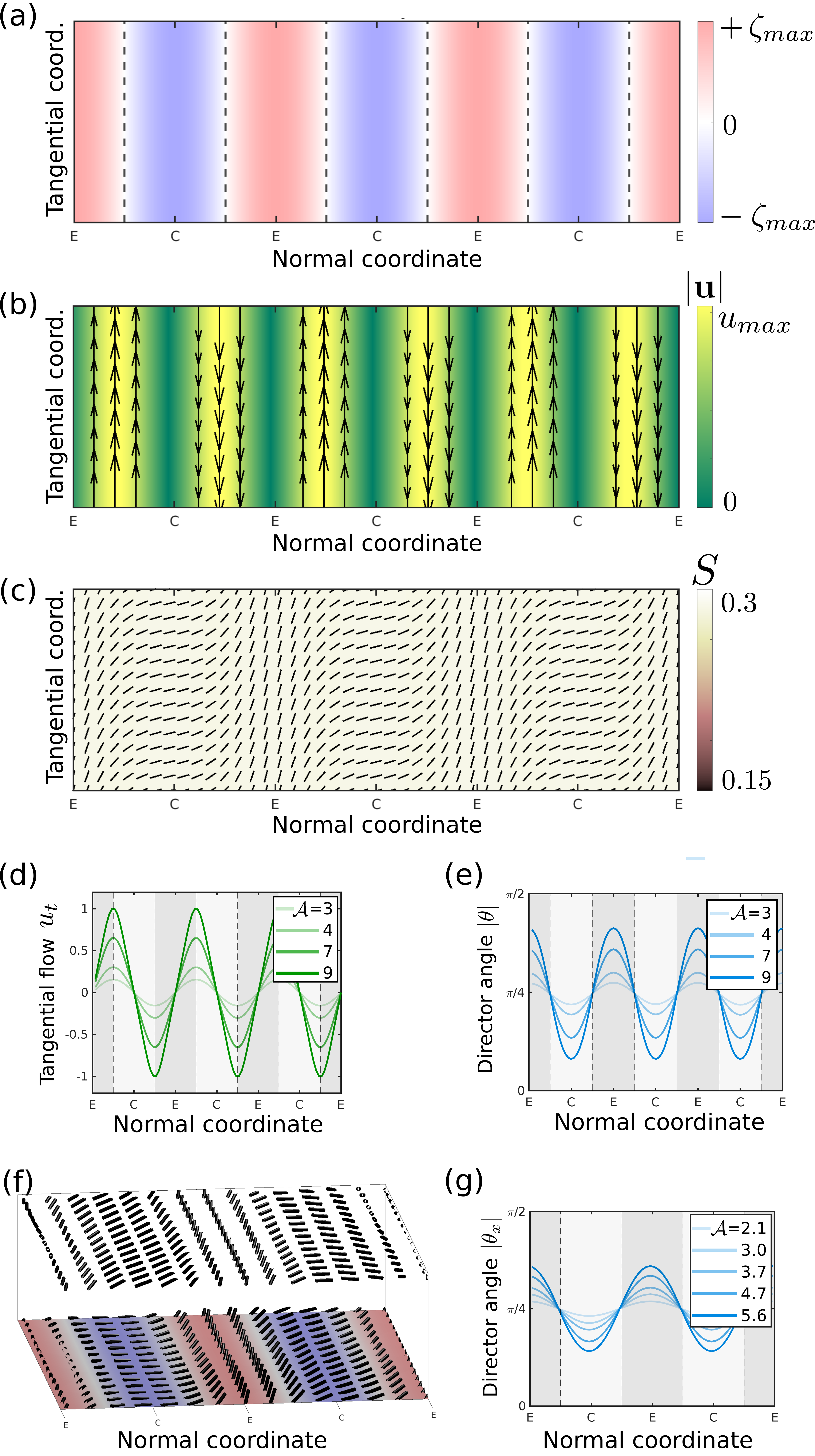}
	\caption{(a) Cosinusoidal variation of active stress along the normal direction. Extensile and contractile regions are labeled as E and C respectively. (b) Active flows form along the tangential direction in the transition regions where $\zeta \approx 0$. (c) Director field aligns tangential (normal) to the gradient direction $\mathbf{k}$ in extensile (contractile) regions. (d,e) Before the onset of active turbulence, the magnitude of flows and the degree of tangential (normal) alignment increases with activity number $\mathcal{A}=\lambda \sqrt{|\zeta_{max}|/K_{el}}$. (f,g) Three-dimensional systems in which activity varies along one coordinate axis show the same quasi-two-dimensional behaviour. However, the director alignment in extensile regions is degenerate within the two-dimensional tangential plane.\\
	}
	\label{fig:1}
\end{figure}

We start by summarising results for systems with heterogeneous activity  in which the magnitude of active stress is too small to create active defects. Assuming gradients in the order parameter $\mathbf{Q}$ are small, the active force density can be approximated by
\begin{equation}
     \mathbf{F}_{act} \approx S \left( \mathbf{\nabla} \zeta - 2 \left( \mathbf{\nabla} \zeta \cdot \mathbf{n} \right) \mathbf{n} \right) \: , 
\end{equation}
This approximation is justified as long as $|\mathbf{\nabla} \zeta| \gg |\zeta ~ \mathbf{n}(\mathbf{\nabla} \cdot \mathbf{n})|$, which is satisfied in the transition area between extensile and contractile domains $\left(\zeta \approx 0 \right)$ or in the absence of topological defects in active regions which create strong distortions in the order parameter. The force component along the tangential direction, perpendicular to the activity gradient, is \citep{bhattacharyya2021coupling}
\begin{equation}
    F^{||}_{act} = - 2 |\mathbf{\nabla} \zeta| \cdot S \cos\theta \sin\theta \: ,
\end{equation}
where $\theta \in [-\pi/2, \pi/2]$ denotes the angle of director field $\mathbf{n}$ with respect to the activity gradient vector $\mathbf{\nabla} \zeta$. Whenever $\mathbf{n}$ is not aligned parallel or perpendicular to a finite activity gradient, active forces will generate flows along the tangential direction, which are maximal at the steepest point of the activity gradient (Fig.~\ref{fig:1} b,d). As a consequence of the co-rotation term $\mathbf{\mathcal{W}}$ in eq.~(\ref{eq:Q}), the resulting shear flow rotates the director field $\mathbf{n}$ until a stable equilibrium is reached where the elastic energy associated with director deformations balances co-rotation by the flow. 

In the case of $0< \theta < \pi/2$ and considering a $\mathbf{\nabla} \zeta > 0$, the resulting shear flow rotates the director clockwise left of the maximum of $|\mathbf{\nabla} \zeta|$ and anti-clockwise to its right. Hence the director in the contractile domain points along the normal, parallel to the activity gradient $\mathbf{k}$, while in the extensile domain the director aligns tangentially, perpendicular to the activity gradient (Fig.~\ref{fig:1} c). The magnitude of nematic alignment depends on the relative strengths of the flow-induced rotation of the director field through the co-rotational term $\mathcal{W}$ and the relaxation of nematic distortions through the molecular field $\mathbf{H}$. Thus spatial variations of $|\theta|$ along the normal coordinate become larger for increasing activity $\zeta_{max}$ (faster flow) and decreasing elastic constant $K_{el}$ (slower relaxation) (Fig.~\ref{fig:1} e). 

The nematic alignment induced by active flows at points of large activity gradients therefore acts as an effective anchoring force, favouring normal anchoring on the contractile side and tangential anchoring on the extensile side of transition regions. To investigate how the shape of activity profiles modifies the dynamics, we also used activity profiles which feature sharper interfaces between extensile and contractile domains, 
\begin{equation}
    \zeta(\mathbf{x}) = 
    \begin{cases}
    \zeta_{max} ,& \text{if } 0 \leq x\leq L/3,\\
    -\zeta_{max} ,& \text{if } L/2 \leq x\leq 5/6 L,\\
    \zeta_{max} \cdot \cos\left(\frac{6 \pi}{L} x \right) ,& \text{otherwise}.
    \end{cases}
    \: 
\end{equation}
We find that this does not qualitatively change the dynamics in the laminar regime and only slightly modifies the shape of flow and director profiles (Fig.~S1 a,b \citep{si}).

In 3D systems where activity varies along one of the coordinate axes, the director alignment with respect to the gradient direction $\mathbf{k}$ is quantified by the director angle $|\theta_x| = |\text{cos}^{-1}(\mathbf{k}/|\mathbf{k}| \cdot \mathbf{n})|$. As for the 2D case, the director field follows variations in activity and aligns normal to the activity gradient in contractile domains. In extensile domains, however, the tangential director orientation is degenerate and points along a random direction within the two-dimensional plane perpendicular to $\mathbf{k}$ (Fig.~\ref{fig:1} f,g). 

\section{\label{subsec:2d} Defect polarization in 2D active turbulence}

\begin{figure}
	\centering
	\includegraphics[width=8.7cm]{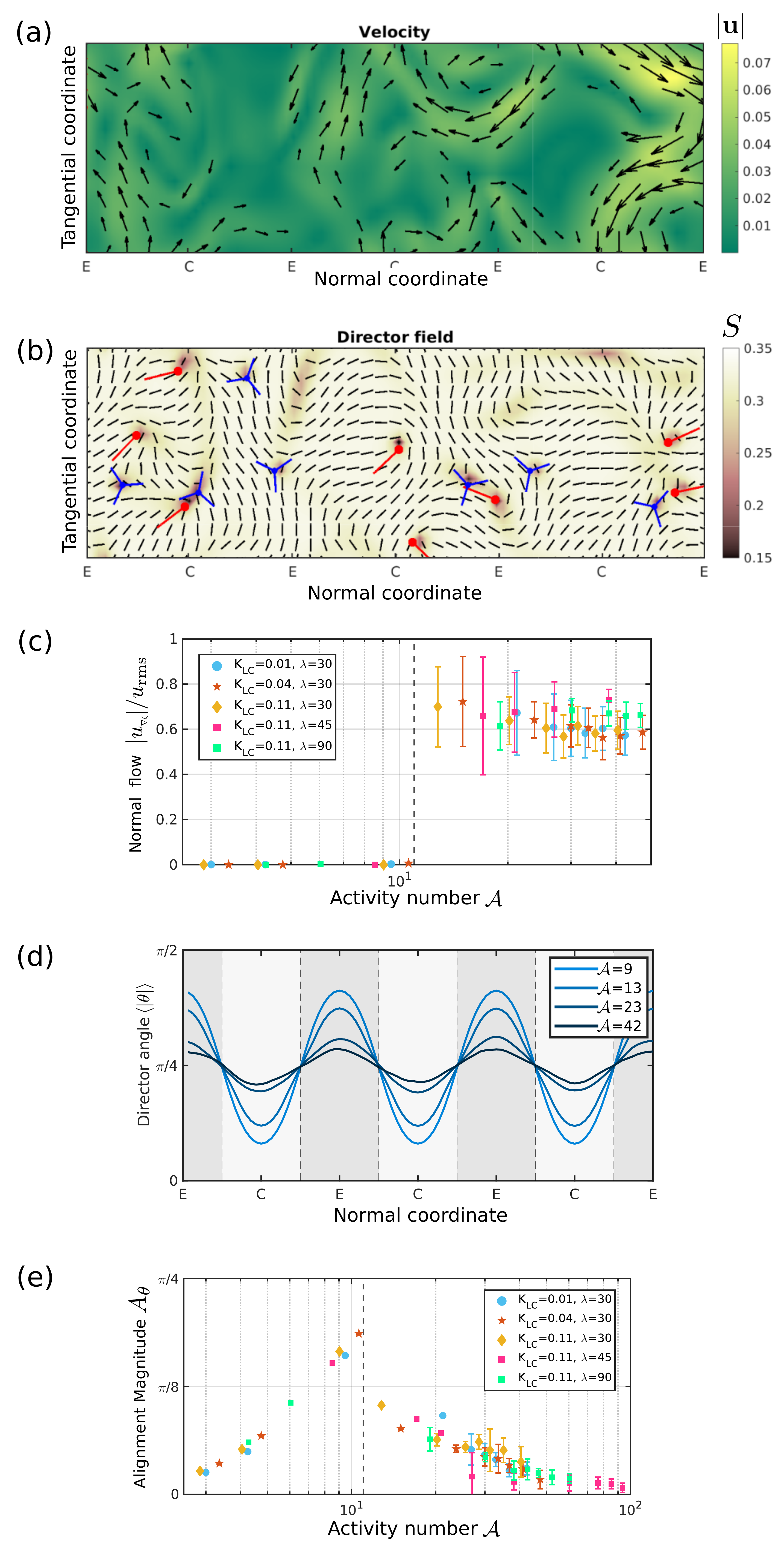}
	\caption{For sufficiently large activity, the system transitions into an active turbulent state, which is characterised by (a) chaotic flow fields and (b) motile topological defects. Comet-like $+1/2$ defects are self-propelled and are shown in red; trefoil-like $-1/2$ defects are only passively advected and are highlighted in blue. (c) The onset of active turbulence is characterised by chaotic flows with components along the normal coordinate, as indicated by a sharp increase of the average flow component along the normal coordinate $\mathbf{u}_N$ relative to the root-mean-squared velocity $\mathbf{u}_{rms}$ in the system. The critical activity $\zeta_{max}$ at which this transition occurs depends on the dimensionless activity number $\mathcal{A}=\lambda \sqrt{|\zeta_{max}|/K_{el}} \sim 10$. (d) The average alignment of the director field in the turbulent regime decreases with $\mathcal{A}$ as shown by the time-averaged director angle $\langle |\theta|  \rangle = \langle |\text{cos}^{-1}(\mathbf{k}/|\mathbf{k}| \cdot \mathbf{n})| \rangle$. (e) The magnitude of alignment $A_\theta$, which is defined as the amplitude of variations of $\langle |\theta|  \rangle$ along the normal direction, increases monotonically with $\mathcal{A}$ in the laminar regime and decays with increasing activity in the turbulent regime. The mean director angle and normal flow component were calculated by averaging quantities over the tangential coordinate and over time $t=[0,80000]$ ($80$ snapshots).
}
	\label{fig:2}
\end{figure}

We will now consider systems with large active stress $\zeta_{max}$, where the soft confinement introduced by passive boundaries between extensile and contractile regions is too weak to prevent the system transitioning to active turbulence. Active turbulence is characterised by strong vorticity and topological $\pm 1/2$ defects which are continually created and destroyed (Fig.~\ref{fig:2} a,b). In 2D systems the transition occurs at a critical activity number $\mathcal{A} \gtrsim 10$, thus depending on the magnitude of active stress $\zeta_{max}$, the elastic constant $K_{el}$ and the wavelength of the activity patterns (Fig.~\ref{fig:2} c). 
Although topological defects and chaotic flow fields tend to destroy any nematic ordering, the time-averaged director field still shows some finite normal (tangential) alignment in contractile (extensile) domains in the turbulent state as shown by the time-averaged director angle $\langle |\theta|  \rangle$ between director $\mathbf{n}$ and gradient axis (Fig.~\ref{fig:2} d). The magnitude of alignment $A_\theta$, which is defined as the amplitude of variations of $\langle |\theta|  \rangle$ along the normal direction, reaches a maximum at the onset of active turbulence and decays with increasing activity (Fig.~\ref{fig:2} e). The director $\mathbf{n}$ eventually approaches a random distribution, $\theta \sim \mathcal{U} \left( -\pi/2,\pi/2 \right)$, which yields $\langle |\theta| \rangle \rightarrow \pi/4$ throughout the system.

\begin{figure}
	\centering
	\includegraphics[width=8.7cm]{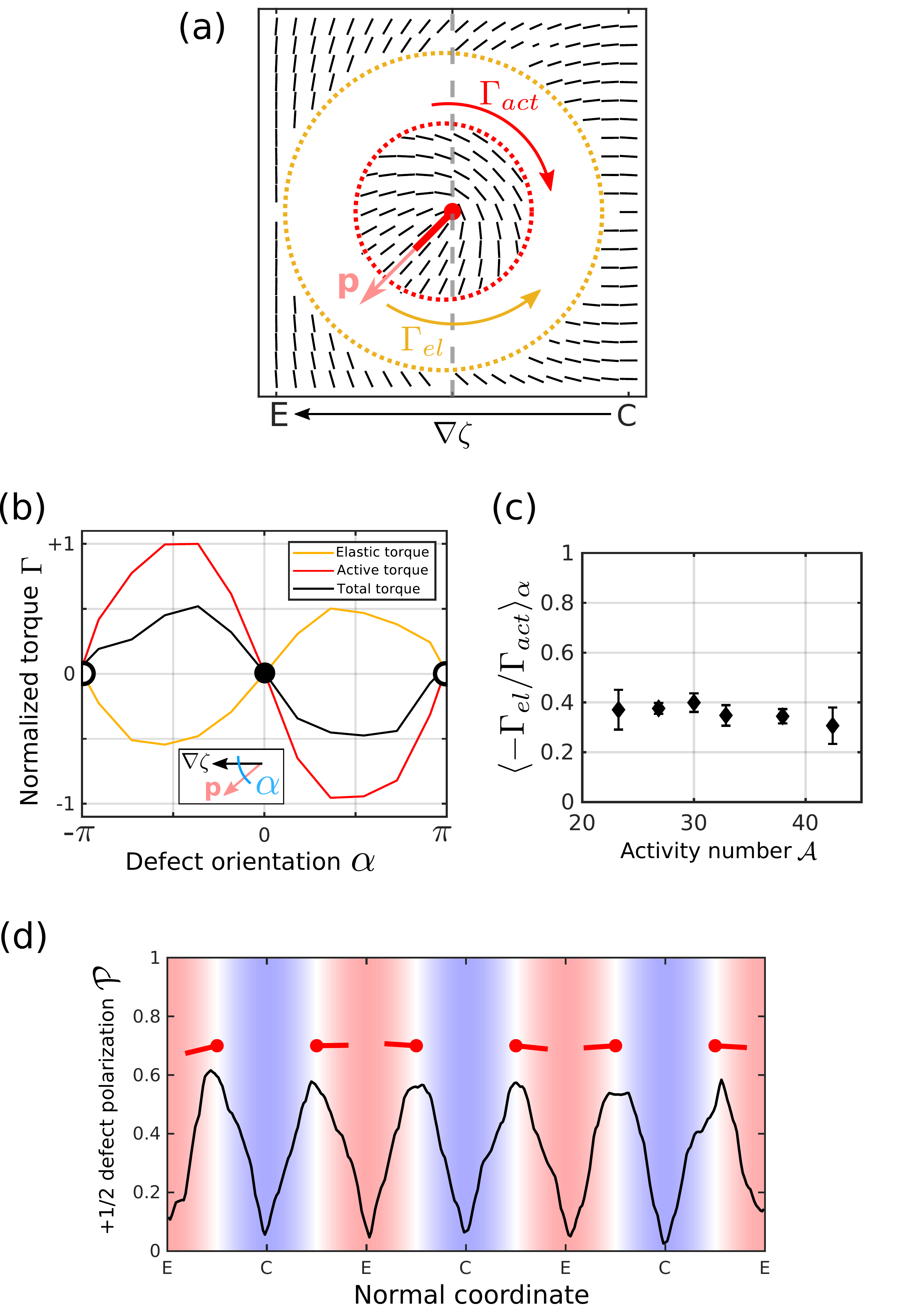}
	\caption{(a) $+1/2$ defects are subject to an elastic torque $\Gamma_{el}$, which tries to minimize the elastic energy of the system (orange), and an active torque $\Gamma_{act}$, which is caused by differential self-propulsion $\mathbf{v}$ across the defect (red). (b) Activity gradients create an active torque on $+1/2$ defects, which is strongest if the defect orientation $\mathbf{p}$ is perpendicular to the activity gradient $\mathbf{\nabla} \zeta$ ($\alpha = \pm \pi/2$) and vanishes at $\alpha = 0,\pi$. Active torques align $+1/2$ defects parallel to activity gradients, which is evident from the stable fixed-point at $\alpha = 0$. Elastic torques, on the other hand, oppose active torques and favour $+1/2$ defects to align anti-parallel to activity gradients (see Fig.~S3 \citep{si}). We find that the average torque on defects $\Gamma = \Gamma_{el} + \Gamma_{act}$ is dominated by active contributions in the regime of active turbulence. Average torques $\Gamma_{act}$ and $\Gamma_{el}$ were obtained by averaging over defects within an orientation window $[\alpha-\pi/12, \alpha+\pi/12]$. (c) The relative magnitude of active and elastic torques does not significantly change as function of activity number $\mathcal{A}$. Error bars show the mean and standard deviation of $\Gamma_{el}/\Gamma_{act}$ when averaged over defect orientations $\alpha$. (d) $+1/2$ defects are polarized at positions of large activity gradients between contractile and extensile domains with their tail pointing towards extensile domains. The fluctuations in polarization $\mathcal{P}$ are caused by the finite sample size of defect orientations. The defect polarization $\mathcal{P}$ as a function of normal coordinate $x$ was obtained by averaging over defect orientations $\mathbf{p}$ in intervals $[x-\lambda/20, x+\lambda/20]$ along  $x$ and averaging over time $t=[0,400000]$ ($200$ snapshots).}
	\label{fig:3}
\end{figure}

In the active turbulent regime, both $\pm 1/2$ defects move in the background flow, but comet-shaped $+1/2$ defects are also self-motile, with their propulsion speed being proportional to activity $|\zeta|$ \citep{giomi2015geometry}. Their propulsion direction $\mathbf{v}$ is parallel (anti-parallel) to the defect orientation vector $\mathbf{p} = \mathbf{\nabla} \cdot \mathbf{Q} / |\mathbf{\nabla} \cdot \mathbf{Q}|$ (which points from head to tail), in contractile (extensile) systems. Activity gradients across $+1/2$ defects give rise an active torque $\Gamma_{act}$ which acts to align $+1/2$ defects parallel to activity gradients, $\mathbf{p} \propto \mathbf{\nabla} \zeta$ (Fig.~S3 a,b \citep{si}). On the other hand, the flow-driven normal (tangential) director field alignment in contractile (extensile) domains acts as a soft boundary condition on the director field $\mathbf{n}$. $+1/2$ defects which align anti-parallel to the activity gradient, $\mathbf{p} \propto -\mathbf{\nabla} \zeta$, minimize distortions in $\mathbf{n}$, thus elastic forces create an elastic torque $\Gamma_{el}$ which minimize the elastic energy of the system (Fig.~S3 c,d \citep{si}). The elastic torque $\Gamma_{el}$ on $+1/2$ defects thus opposes the active torque $\Gamma_{act}$ (Fig.~\ref{fig:3} a). While the magnitude of active and elastic torques can be easily measured for single $+1/2$ defects subject to external activity gradients or nematic boundary conditions, it is not immediately clear which contribution dominates in the regime of active turbulence, where a finite alignment magnitude $A_\theta$ of the average director field $\langle |\theta| \rangle$ gives rise to soft boundary conditions and interactions between defects become important.

To investigate the behaviour of defects in the regime of active turbulence, we track the position and orientation $\mathbf{p}$ of $+ 1/2$ defects in the system. By considering active and elastic forces in the vicinity of defects, we can calculate elastic and active torques on $+1/2$ defects as a function of defect orientation (see Supplementary Information \citep{si}). By averaging over defects and time, we confirm that elastic torques indeed oppose active torques (Fig.~\ref{fig:3} b). However, active torques remain the dominant contribution to the total torque $\Gamma = \Gamma_{el} + \Gamma_{act}$ in the regime of active turbulence (Fig.~\ref{fig:3} c).

In order to quantify to which extent defects are locally polarized by active torques, we measure the angular distribution of defect orientations $\mathbf{p}$, averaging over time (Fig.~S2 a \citep{si}). We define the defect orientation order parameter $\mathcal{P} = \sqrt{ \langle p_x \rangle^2 + \langle p_y \rangle^2 }$, which is is zero if $+1/2$ defects are oriented randomly and equals one if the defects perfectly align along one direction. The polarization direction, $\langle \mathbf{p} \rangle$, is calculated as the circular mean of the distribution of $\mathbf{p}$. Even though the motion of individual defects is highly chaotic in the regime of active turbulence, we find that $+1/2$ defects are strongly polarized at the boundaries between extensile and contractile domains (Fig.~\ref{fig:3} d). The average defect orientation aligns with and is proportional to the activity gradient, $\langle \mathbf{p} \rangle \propto \mathbf{\nabla} \zeta$, in agreement with previous theoretical predictions which considered active torques on individual defects \citep{zhang2021spatiotemporal, shankar2019hydrodynamics}. A strong polarization of defects along activity gradients thus confirms once more that active torques on $+1/2$ defects are dominating elastic torques, $\Gamma_{act} > \Gamma_{el}$. It should be emphasised, however, that both torques originate from activity gradients since the nematic boundary condition giving rise to $\Gamma_{el}$ is driven by active tangential flows.

In systems with sharp activity interfaces, non-zero defect polarization $\langle \mathbf{p} \rangle$ and average director alignment $\langle |\theta| \rangle$ are limited to boundaries and vanish in contractile and extensile bulk regions as expected for isotropic active turbulence (Fig.~S1 c,d \citep{si}).

\section{\label{subsec:3d} Disclination line properties in 3D active turbulence }

 \begin{figure*}
	\includegraphics[width=17cm]{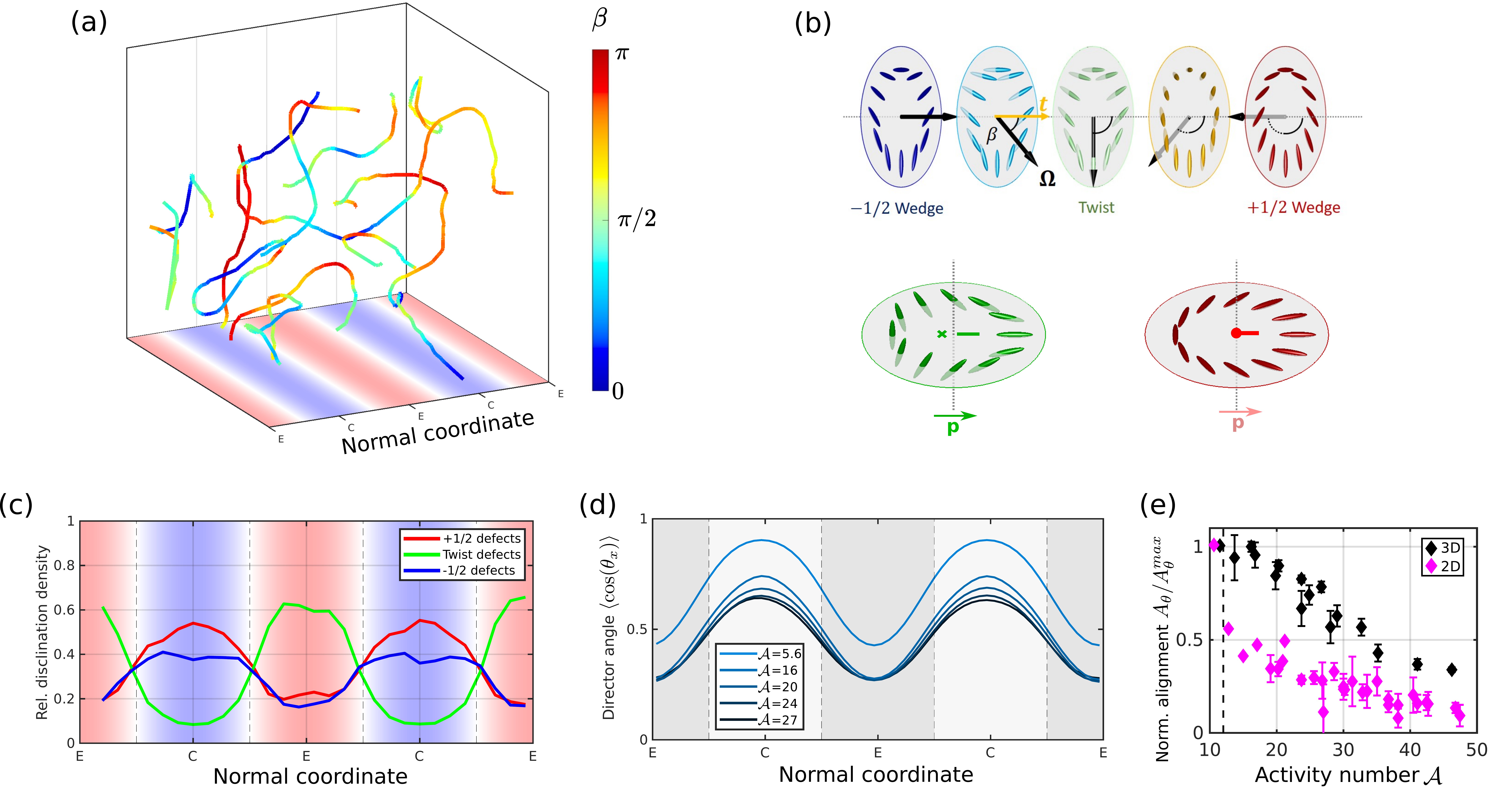}
	\caption{(a) Three-dimensional active turbulence is characterised by motile disclination lines which form closed loops or span the entire system as open lines. (b) Disclination lines can be locally classified by a continuous twist-angle $0 < \beta < \pi$ which quantifies how the director field winds around the disclination core in the plane perpendicular to the line tangent. $\beta$ is the angle between the line tangent $\mathbf{t}$ and the rotation axis $\mathbf{\Omega}$ around which the director winds while moving around the disclination core. Quasi-two-dimensional $\pm 1/2$ defects correspond to $\beta = \approx 0, \pi$, respectively, and $\beta \approx \pi/2$ describes twist disclinations, which are unique to three-dimensional systems (radial twist-type shown here). (c) The relative density of $\pm 1/2$ and twist segments depends on the type of activity (extensile/contractile) and follows spatial variations in active stress (fluctuations are caused by finite sample size). (d) The director field alignment in 3D systems in the turbulent regime decays with $\mathcal{A}=\lambda \sqrt{|\zeta_{max}|/K_{el}}$, as shown by the time-averaged director angle $\langle \cos(\theta_x)  \rangle$. (e) As for 2D systems, the magnitude of alignment $A_\theta$, which in 3D is defined as the amplitude of variations of $\langle \cos(\theta_x) \rangle$ along the normal direction, decays with increasing activity number $\mathcal{A}$. Black diamonds show $A_\theta/A_\theta^{max}$ as a function of $\zeta_{max}$ for different parameters $K_{el} = [0.04,0.08,0.12,0.2]$, where $A_\theta^{max}$ is the maximum alignment realized just before the onset of active turbulence (dashed line). Pink diamonds show normalized data taken from Fig.~\ref{fig:2} e. The mean director angle and disclination line densities were calculated by averaging quantities over both tangential coordinates and time $t=[0,500000]$ ($250$ snapshots).}
	\label{fig:4}
\end{figure*}

We next consider 3D bulk systems in which activity varies along one direction. 
The point defects of 2D nematics are replaced by disclination lines which can continuously transform from a local $-1/2$ configuration, in the plane perpendicular to the line tangent, into a $+1/2$ configuration through an intermediate twist defect, which is inherently three-dimensional (Fig.~\ref{fig:4} a,b). The director configuration of the defects along a disclination line can be locally classified by the {\it twist angle} $\beta$, where $-1/2$ ($+1/2$) wedge-type defects correspond to $\beta=0$ $(\pi)$ and line segments with twist defects are indicated by $\beta = \pi/2$. Radial twist and $+1/2$ defects both have a unique polar direction $\mathbf{p}$ in the plane perpendicular to the line tangent (Fig.~\ref{fig:4} b). 

The flows and morphological dynamics of disclination lines in 3D active nematics are governed by the local director profile \citep{binysh2020three, vcopar2019topology}. While disclination lines in purely extensile systems are predominantly twist-like, contractile activity favours the creation of the wedge-like line segments with cross-sections that resemble $\pm 1/2$ defects \citep{nejad2022active, ruske2021morphology}. This is reflected by  variations in the spatial density of $\pm 1/2$ and twist line segments across a system with extensile and contractile domains (Fig.~\ref{fig:4} c).

In 3D active turbulence with homogeneous activity the orientation of the director field $\mathbf{n}$ is, by symmetry, randomly distributed, $\cos \theta_x \sim \mathcal{U}(0,1)$, where $\theta_x$ is the angle between the director field and a random reference vector. Deviations from the average director angle $\langle \cos(\theta_x) \rangle = 1/2$ thus indicate that the time-averaged director field is either aligned parallel $\left( \langle \cos \theta_x \rangle > 1/2 \right)$ or perpendicular $\left( \langle \cos \theta_x \rangle < 1/2 \right)$ with respect to the reference vector. For 3D active turbulence with heterogeneous activity we find that the time-averaged director field is aligned normal (tangential) with respect to the activity gradient in contractile (extensile) domains, as for 2D systems (Fig.~\ref{fig:4} d). The nematic alignment in 3D systems can be quantified by $A_\theta$, the amplitude of variations of $\langle \cos(\theta_x) \rangle$ along the normal direction. To compare the variation of the alignment magnitude with activity in 2D ($A_\theta \in [0,\pi/4]$) and 3D systems ($A_\theta \in [0,0.5]$), we normalize $A_\theta$ with respect to $A_\theta^{max}$, the maximum alignment realized just before the onset of active turbulence. We find that with increasing activity, the average alignment of the director field in 3D decays much {more slowly} compared to 2D systems, where $A_\theta$ sharply drops at the onset of active turbulence (Fig.~\ref{fig:4} e).

\begin{figure*}
	\includegraphics[width=17cm]{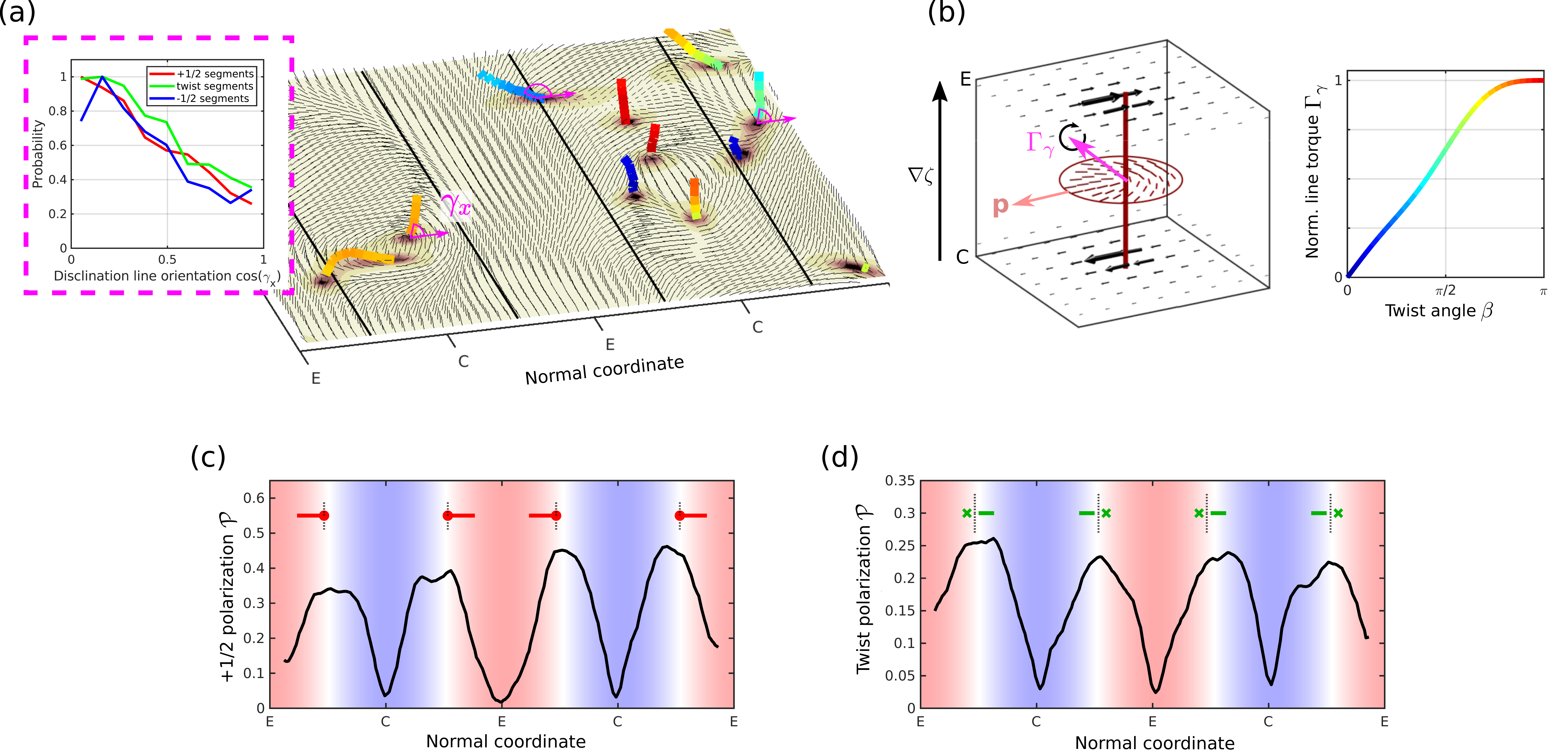}
	\caption{(a) The angle $\gamma_x$ between disclination line segments and the normal axis along which activity varies is measured for all disclination lines in the system. The time-averaged distribution of $|\cos(\gamma_x)|$ shows that disclination line segments preferentially point perpendicular to activity gradients ($\gamma_x = \pi/2$), independent of twist-angle $\beta$. The distribution of orientation $\gamma_x$ was obtained by averaging over all disclination line segments and time $t=[0,500000]$. (b) When a $+1/2$ disclination line points along an activity gradients $\mathbf{\nabla} \zeta$ ($\gamma_x=0$), active forces (black arrows) create a torque $\Gamma_\gamma$ (pink arrow), which re-orients the disclination line perpendicular the gradient direction ($\gamma_x = \pi/2$). The magnitude of the line torque $\Gamma_\gamma$ depends on the self-propulsion speed of the disclination type, thus it is strongest for $+1/2$ line segments and vanishes for $-1/2$ segments. Since disclination line segments of different types $\beta$ are connected with each other, line torques $\Gamma_\gamma$ acting on $+1/2$ and twist segments also align passive $-1/2$ segments. (c) As in 2D systems, active torques align the orientation $\mathbf{p}$ of $+1/2$ defects parallel to activity gradients $\mathbf{\nabla} \zeta$, causing a large polarization of $+1/2$ disclinations ($\beta\in[7\pi/8,\pi]$) at points of large activity gradients. (d) Twist disclinations ($\beta\in[3\pi/8,5\pi/8]$) are dominated by elastic torque rather than active torque, thus they are polarized anti-parallel to activity gradients. The fluctuations in polarization $\mathcal{P}$ are caused by the finite sample size of $+1/2$ and twist defect orientations. The polarization $\mathcal{P}$ of disclination lines was calculated by averaging over defect orientations $\mathbf{p}$ in intervals $[x-\lambda/20, x+\lambda/20]$ along the normal coordinate $x$ and averaging over time $t=[0,500000]$ ($250$ snapshots).
	}
	\label{fig:5}
\end{figure*}

To gain further insight, we measure the angle $\gamma_x$ between disclination line segments and the gradient direction $\mathbf{k}$. We find that disclination lines are preferentially oriented perpendicular to activity gradients, $\gamma_x \approx \pi/2$ (Fig.~\ref{fig:5} a). This preferred alignment is caused by active torques $\Gamma_\gamma$ on line segments when they lie along activity gradients (Fig.~\ref{fig:5} b). This implies that the orientation of the defect structures in the plane perpendicular to line segments can be either parallel ($\mathbf{p} \propto \mathbf{\nabla} \zeta$), anti-parallel ($\mathbf{p} \propto -\mathbf{\nabla} \zeta$) or perpendicular ($\mathbf{p} \cdot \mathbf{\nabla} \zeta = 0$) to the activity gradient. 

We define the polarization order parameter for disclination lines as $\mathcal{P} = \sqrt{ \langle p_x \rangle^2 + \langle p_y \rangle^2 + \langle p_z \rangle^2}$, which is is zero if defects are oriented randomly and equals one if the defects perfectly align along one direction. The polarization order parameter $\mathcal{P}$ and direction of $+1/2$ and twist segments are shown in Fig.~\ref{fig:5} c  and Fig.~\ref{fig:5} d, respectively. While $+1/2$ disclination are polarized parallel to activity gradients, $\langle \mathbf{p} \rangle \propto \mathbf{\nabla} \zeta$, like in 2D systems, twist disclinations are polarized anti-parallel to the activity gradient $\langle \mathbf{p} \rangle \propto -\mathbf{\nabla} \zeta$.  

We argue that this is caused by a different balance between elastic torques $\Gamma_{el}$ and active torques $\Gamma_{act}$ on different defect types. While active torques $\Gamma_{act}$ dominate elastic torques $\Gamma_{el}$ for $+1/2$ disclinations (see Fig.~\ref{fig:3} b), active forces around twist disclinations feature out-of-plane components, yielding an active torque which does not align with the line tangent (Fig.~S6 \citep{si}). Thus the orientation $\mathbf{p}$ of twist defects is less affected by active torques than the orientation of $+1/2$ defects (Fig.~S6 c), causing twist defects to align in a way which minimizes the elastic energy of the system. Indeed, since active torque is generated by differential velocities across defects, we would expect a smaller active torque on twist defects because they have a lower self-propulsion speed than $+1/2$ defects \citep{binysh2020three, ruske2021morphology}. 

Sharp activity interfaces in 3D systems give rise to similar dynamics, where defect polarization $\langle \mathbf{p} \rangle$ and director alignment $\langle \cos \theta_x \rangle$ are limited to boundary regions and vanish in contractile and extensile bulk regions (Fig.~S4 \citep{si}).

\section{\label{sec:conclusion} Conclusion}

In this paper we have investigated how two- and three-dimensional active nematic bulk systems respond to one-dimensional spatial variations of active stress. We have shown that, in the absence of defects, activity gradients induce an effective anchoring force which aligns the director field along the normal direction in contractile domains and tangentially in extensile domains. In line with recent analytical work \citep{shankar2019hydrodynamics, tang2021alignment}, we found that the orientation of motile $+1/2$ defects in 2D active turbulence are dominated by active torque and polarized parallel to activity gradients, $\mathbf{p}\propto \mathbf{\nabla}\zeta$. Since the director field in the vicinity of polarized $+1/2$ defects opposes the director alignment set up by active flows perpendicular to activity gradients, the average alignment of the director field in 2D active turbulence remains weak.

In 3D systems we find that disclination line segments are oriented perpendicular to activity gradients, thereby allowing the orientation of the defect structures in the plane perpendicular to line segments to point parallel, anti-parallel or perpendicular to activity gradients. Like in 2D systems, active torque robustly aligns $+1/2$ disclination lines parallel to activity gradients, even in the turbulent regime. In contrast, the alignment of twist disclinations is dominated by elastic interactions rather than active torques, causing them to be polarized anti-parallel to activity gradients. Thereby the director field in the vicinity of twist defects matches with the alignment set up by active flows in the plane perpendicular to activity gradients and the average alignment of the director field decays more slowly with activity number $\mathcal{A}$ than in 2D systems. Deriving exact expressions for the elastic torque on disclination line segments in the turbulent regime remains challenging since the local torque on line segments depends not only on the nematic boundary conditions in the plane perpendicular to the line segment, but also on elastic forces acting along disclination lines and their local curvature. Active-passive systems which are purely extensile or contractile with spatial activity profiles $\zeta(\mathbf{x}) = \zeta_{max} (\cos(\mathbf{k} \cdot \mathbf{x}) \pm 1)$ are qualitatively similar to extensile-contractile systems and are presented in the Supplementary Information \citep{si} (Fig.~S2,S5). 

We note that the alignment of the director field and defects due to active force gradients is not unique to bulk systems. In systems with isotropic-nematic interfaces ($\mathbf{\nabla} S > 0$), active forces create flows which align the director field either perpendicular or planar to the interface in contractile or extensile systems, respectively \citep{blow2014biphasic}. This effect, termed \textit{active anchoring}, also affects the orientation of defects close to an interface and can be observed both in experiments \citep{dell2018growing} and in simulations in two and three dimensions \citep{doostmohammadi2016defect, ruske2021morphology}.

Experiments with spatially structured activity have been realised by combining actin filaments with light-activated gear-shifting myosin motors, where light locally induces extensile active stress in an otherwise passive fluid \citep{zhang2021spatiotemporal}. Extending the concept of light-modulated activity to contractile systems, it might be possible to create artificial systems with distinct extensile and contractile domains in the future by selectively activating molecular motors mediating either extensile or contractile stress. Spatial variations of active stress are also ubiquitous in biological systems, ranging from structure formation during embryonic development driven by a heterogeneous distribution of myosin motors \citep{streichan2018global,he2014apical,mason2013apical} to flows and deformations of cell aggregates, such as tumors, caused by differential cell growth and death as a consequence of limited access to nutrients in the core of the aggregate \citep{jagiella2016inferring, delarue2013mechanical, delarue2014stress, ranft2010fluidization}. Understanding the response of two- and three-dimensional active nematic systems to activity gradients is thus an important step towards applying physical theories in the life sciences.

\section*{Conflicts of interest}
There are no conflicts to declare.

\section*{Acknowledgements}
This project was funded by the European Commission’s Horizon 2020 research and innovation programme under the Marie Sklodowska-Curie grant agreement No 812780.

\providecommand*{\mcitethebibliography}{\thebibliography}
\csname @ifundefined\endcsname{endmcitethebibliography}
{\let\endmcitethebibliography\endthebibliography}{}


\begin{mcitethebibliography}{39}
\providecommand*{\natexlab}[1]{#1}
\providecommand*{\mciteSetBstSublistMode}[1]{}
\providecommand*{\mciteSetBstMaxWidthForm}[2]{}
\providecommand*{\mciteBstWouldAddEndPuncttrue}
  {\def\EndOfBibitem{\unskip.}}
\providecommand*{\mciteBstWouldAddEndPunctfalse}
  {\let\EndOfBibitem\relax}
\providecommand*{\mciteSetBstMidEndSepPunct}[3]{}
\providecommand*{\mciteSetBstSublistLabelBeginEnd}[3]{}
\providecommand*{\EndOfBibitem}{}
\mciteSetBstSublistMode{f}
\mciteSetBstMaxWidthForm{subitem}
{(\emph{\alph{mcitesubitemcount}})}
\mciteSetBstSublistLabelBeginEnd{\mcitemaxwidthsubitemform\space}
{\relax}{\relax}

\bibitem[J{\"u}licher \emph{et~al.}(2018)J{\"u}licher, Grill, and
  Salbreux]{julicher2018hydrodynamic}
F.~J{\"u}licher, S.~W. Grill and G.~Salbreux, \emph{Reports on Progress in
  Physics}, 2018, \textbf{81}, 076601\relax
\mciteBstWouldAddEndPuncttrue
\mciteSetBstMidEndSepPunct{\mcitedefaultmidpunct}
{\mcitedefaultendpunct}{\mcitedefaultseppunct}\relax
\EndOfBibitem
\bibitem[Marchetti \emph{et~al.}(2013)Marchetti, Joanny, Ramaswamy, Liverpool,
  Prost, Rao, and Simha]{marchetti2013hydrodynamics}
M.~C. Marchetti, J.-F. Joanny, S.~Ramaswamy, T.~B. Liverpool, J.~Prost, M.~Rao
  and R.~A. Simha, \emph{Reviews of Modern Physics}, 2013, \textbf{85},
  1143\relax
\mciteBstWouldAddEndPuncttrue
\mciteSetBstMidEndSepPunct{\mcitedefaultmidpunct}
{\mcitedefaultendpunct}{\mcitedefaultseppunct}\relax
\EndOfBibitem
\bibitem[Chen \emph{et~al.}(2018)Chen, Saw, M{\`e}ge, and
  Ladoux]{chen2018mechanical}
T.~Chen, T.~B. Saw, R.-M. M{\`e}ge and B.~Ladoux, \emph{Journal of Cell
  Science}, 2018, \textbf{131}, jcs218156\relax
\mciteBstWouldAddEndPuncttrue
\mciteSetBstMidEndSepPunct{\mcitedefaultmidpunct}
{\mcitedefaultendpunct}{\mcitedefaultseppunct}\relax
\EndOfBibitem
\bibitem[Lee \emph{et~al.}(2021)Lee, Leech, Rust, Das, McGorty, Ross, and
  Robertson-Anderson]{lee2021myosin}
G.~Lee, G.~Leech, M.~J. Rust, M.~Das, R.~J. McGorty, J.~L. Ross and R.~M.
  Robertson-Anderson, \emph{Science Advances}, 2021, \textbf{7}, eabe4334\relax
\mciteBstWouldAddEndPuncttrue
\mciteSetBstMidEndSepPunct{\mcitedefaultmidpunct}
{\mcitedefaultendpunct}{\mcitedefaultseppunct}\relax
\EndOfBibitem
\bibitem[Yaman \emph{et~al.}(2019)Yaman, Demir, Vetter, and
  Kocabas]{yaman2019emergence}
Y.~I. Yaman, E.~Demir, R.~Vetter and A.~Kocabas, \emph{Nature Communications},
  2019, \textbf{10}, 1--9\relax
\mciteBstWouldAddEndPuncttrue
\mciteSetBstMidEndSepPunct{\mcitedefaultmidpunct}
{\mcitedefaultendpunct}{\mcitedefaultseppunct}\relax
\EndOfBibitem
\bibitem[Saw \emph{et~al.}(2017)Saw, Doostmohammadi, Nier, Kocgozlu, Thampi,
  Toyama, Marcq, Lim, Yeomans, and Ladoux]{saw2017topological}
T.~B. Saw, A.~Doostmohammadi, V.~Nier, L.~Kocgozlu, S.~Thampi, Y.~Toyama,
  P.~Marcq, C.~T. Lim, J.~M. Yeomans and B.~Ladoux, \emph{Nature}, 2017,
  \textbf{544}, 212--216\relax
\mciteBstWouldAddEndPuncttrue
\mciteSetBstMidEndSepPunct{\mcitedefaultmidpunct}
{\mcitedefaultendpunct}{\mcitedefaultseppunct}\relax
\EndOfBibitem
\bibitem[Mueller \emph{et~al.}(2019)Mueller, Yeomans, and
  Doostmohammadi]{mueller2019emergence}
R.~Mueller, J.~M. Yeomans and A.~Doostmohammadi, \emph{Physical Review
  Letters}, 2019, \textbf{122}, 048004\relax
\mciteBstWouldAddEndPuncttrue
\mciteSetBstMidEndSepPunct{\mcitedefaultmidpunct}
{\mcitedefaultendpunct}{\mcitedefaultseppunct}\relax
\EndOfBibitem
\bibitem[Duclos \emph{et~al.}(2017)Duclos, Erlenk{\"a}mper, Joanny, and
  Silberzan]{duclos2017topological}
G.~Duclos, C.~Erlenk{\"a}mper, J.-F. Joanny and P.~Silberzan, \emph{Nature
  Physics}, 2017, \textbf{13}, 58--62\relax
\mciteBstWouldAddEndPuncttrue
\mciteSetBstMidEndSepPunct{\mcitedefaultmidpunct}
{\mcitedefaultendpunct}{\mcitedefaultseppunct}\relax
\EndOfBibitem
\bibitem[Saw \emph{et~al.}(2018)Saw, Xi, Ladoux, and Lim]{saw2018biological}
T.~B. Saw, W.~Xi, B.~Ladoux and C.~T. Lim, \emph{Advanced Materials}, 2018,
  \textbf{30}, 1802579\relax
\mciteBstWouldAddEndPuncttrue
\mciteSetBstMidEndSepPunct{\mcitedefaultmidpunct}
{\mcitedefaultendpunct}{\mcitedefaultseppunct}\relax
\EndOfBibitem
\bibitem[Vromans and Giomi(2016)]{vromans2016orientational}
A.~J. Vromans and L.~Giomi, \emph{Soft Matter}, 2016, \textbf{12},
  6490--6495\relax
\mciteBstWouldAddEndPuncttrue
\mciteSetBstMidEndSepPunct{\mcitedefaultmidpunct}
{\mcitedefaultendpunct}{\mcitedefaultseppunct}\relax
\EndOfBibitem
\bibitem[Duclos \emph{et~al.}(2020)Duclos, Adkins, Banerjee, Peterson,
  Varghese, Kolvin, Baskaran, Pelcovits, Powers,
  Baskaran,\emph{et~al.}]{duclos2020topological}
G.~Duclos, R.~Adkins, D.~Banerjee, M.~S. Peterson, M.~Varghese, I.~Kolvin,
  A.~Baskaran, R.~A. Pelcovits, T.~R. Powers, A.~Baskaran \emph{et~al.},
  \emph{Science}, 2020, \textbf{367}, 1120--1124\relax
\mciteBstWouldAddEndPuncttrue
\mciteSetBstMidEndSepPunct{\mcitedefaultmidpunct}
{\mcitedefaultendpunct}{\mcitedefaultseppunct}\relax
\EndOfBibitem
\bibitem[Binysh \emph{et~al.}(2020)Binysh, Kos, {\v{C}}opar, Ravnik, and
  Alexander]{binysh2020three}
J.~Binysh, {\v{Z}}.~Kos, S.~{\v{C}}opar, M.~Ravnik and G.~P. Alexander,
  \emph{Physical Review Letters}, 2020, \textbf{124}, 088001\relax
\mciteBstWouldAddEndPuncttrue
\mciteSetBstMidEndSepPunct{\mcitedefaultmidpunct}
{\mcitedefaultendpunct}{\mcitedefaultseppunct}\relax
\EndOfBibitem
\bibitem[{\v{C}}opar \emph{et~al.}(2019){\v{C}}opar, Aplinc, Kos, {\v{Z}}umer,
  and Ravnik]{vcopar2019topology}
S.~{\v{C}}opar, J.~Aplinc, {\v{Z}}.~Kos, S.~{\v{Z}}umer and M.~Ravnik,
  \emph{Physical Review X}, 2019, \textbf{9}, 031051\relax
\mciteBstWouldAddEndPuncttrue
\mciteSetBstMidEndSepPunct{\mcitedefaultmidpunct}
{\mcitedefaultendpunct}{\mcitedefaultseppunct}\relax
\EndOfBibitem
\bibitem[Ruske and Yeomans(2021)]{ruske2021morphology}
L.~J. Ruske and J.~M. Yeomans, \emph{Physical Review X}, 2021, \textbf{11},
  021001\relax
\mciteBstWouldAddEndPuncttrue
\mciteSetBstMidEndSepPunct{\mcitedefaultmidpunct}
{\mcitedefaultendpunct}{\mcitedefaultseppunct}\relax
\EndOfBibitem
\bibitem[Zhang \emph{et~al.}(2021)Zhang, Redford, Ruijgrok, Kumar, Mozaffari,
  Zemsky, Dinner, Vitelli, Bryant,
  Gardel,\emph{et~al.}]{zhang2021spatiotemporal}
R.~Zhang, S.~A. Redford, P.~V. Ruijgrok, N.~Kumar, A.~Mozaffari, S.~Zemsky,
  A.~R. Dinner, V.~Vitelli, Z.~Bryant, M.~L. Gardel \emph{et~al.}, \emph{Nature
  Materials}, 2021, \textbf{20}, 875--882\relax
\mciteBstWouldAddEndPuncttrue
\mciteSetBstMidEndSepPunct{\mcitedefaultmidpunct}
{\mcitedefaultendpunct}{\mcitedefaultseppunct}\relax
\EndOfBibitem
\bibitem[Shankar and Marchetti(2019)]{shankar2019hydrodynamics}
S.~Shankar and M.~C. Marchetti, \emph{Physical Review X}, 2019, \textbf{9},
  041047\relax
\mciteBstWouldAddEndPuncttrue
\mciteSetBstMidEndSepPunct{\mcitedefaultmidpunct}
{\mcitedefaultendpunct}{\mcitedefaultseppunct}\relax
\EndOfBibitem
\bibitem[Tang and Selinger(2021)]{tang2021alignment}
X.~Tang and J.~V. Selinger, \emph{Physical Review E}, 2021, \textbf{103},
  022703\relax
\mciteBstWouldAddEndPuncttrue
\mciteSetBstMidEndSepPunct{\mcitedefaultmidpunct}
{\mcitedefaultendpunct}{\mcitedefaultseppunct}\relax
\EndOfBibitem
\bibitem[Mozaffari \emph{et~al.}(2021)Mozaffari, Zhang, Atzin, and
  de~Pablo]{mozaffari2021defect}
A.~Mozaffari, R.~Zhang, N.~Atzin and J.~J. de~Pablo, \emph{Physical Review
  Letters}, 2021, \textbf{126}, 227801\relax
\mciteBstWouldAddEndPuncttrue
\mciteSetBstMidEndSepPunct{\mcitedefaultmidpunct}
{\mcitedefaultendpunct}{\mcitedefaultseppunct}\relax
\EndOfBibitem
\bibitem[Marenduzzo \emph{et~al.}(2007)Marenduzzo, Orlandini, and
  Yeomans]{marenduzzo2007hydrodynamics}
D.~Marenduzzo, E.~Orlandini and J.~M. Yeomans, \emph{Physical Review Letters},
  2007, \textbf{98}, 118102\relax
\mciteBstWouldAddEndPuncttrue
\mciteSetBstMidEndSepPunct{\mcitedefaultmidpunct}
{\mcitedefaultendpunct}{\mcitedefaultseppunct}\relax
\EndOfBibitem
\bibitem[Schimming and Vi{\~n}als(2020)]{schimming2020anisotropic}
C.~D. Schimming and J.~Vi{\~n}als, \emph{Physical Review E}, 2020,
  \textbf{102}, 010701\relax
\mciteBstWouldAddEndPuncttrue
\mciteSetBstMidEndSepPunct{\mcitedefaultmidpunct}
{\mcitedefaultendpunct}{\mcitedefaultseppunct}\relax
\EndOfBibitem
\bibitem[Long \emph{et~al.}(2021)Long, Tang, Selinger, and
  Selinger]{long2021geometry}
C.~Long, X.~Tang, R.~L. Selinger and J.~V. Selinger, \emph{Soft Matter}, 2021,
  \textbf{17}, 2265--2278\relax
\mciteBstWouldAddEndPuncttrue
\mciteSetBstMidEndSepPunct{\mcitedefaultmidpunct}
{\mcitedefaultendpunct}{\mcitedefaultseppunct}\relax
\EndOfBibitem
\bibitem[Beris and Edwards(1994)]{beris1994thermodynamics}
A.~N. Beris and B.~J. Edwards, \emph{Thermodynamics of flowing systems with
  internal microstructure}, Oxford University Press, New York ; Oxford,
  1994\relax
\mciteBstWouldAddEndPuncttrue
\mciteSetBstMidEndSepPunct{\mcitedefaultmidpunct}
{\mcitedefaultendpunct}{\mcitedefaultseppunct}\relax
\EndOfBibitem
\bibitem[Norton \emph{et~al.}(2018)Norton, Baskaran, Opathalage, Langeslay,
  Fraden, Baskaran, and Hagan]{norton2018insensitivity}
M.~M. Norton, A.~Baskaran, A.~Opathalage, B.~Langeslay, S.~Fraden, A.~Baskaran
  and M.~F. Hagan, \emph{Physical Review E}, 2018, \textbf{97}, 012702\relax
\mciteBstWouldAddEndPuncttrue
\mciteSetBstMidEndSepPunct{\mcitedefaultmidpunct}
{\mcitedefaultendpunct}{\mcitedefaultseppunct}\relax
\EndOfBibitem
\bibitem[Zhang \emph{et~al.}(2018)Zhang, Kumar, Ross, Gardel, and
  De~Pablo]{zhang2018interplay}
R.~Zhang, N.~Kumar, J.~L. Ross, M.~L. Gardel and J.~J. De~Pablo,
  \emph{Proceedings of the National Academy of Sciences}, 2018, \textbf{115},
  E124--E133\relax
\mciteBstWouldAddEndPuncttrue
\mciteSetBstMidEndSepPunct{\mcitedefaultmidpunct}
{\mcitedefaultendpunct}{\mcitedefaultseppunct}\relax
\EndOfBibitem
\bibitem[si()]{si}
\emph{See Supplementary Information for supplementary
  figures and details on numerical implementation.}\relax
\mciteBstWouldAddEndPunctfalse
\mciteSetBstMidEndSepPunct{\mcitedefaultmidpunct}
{}{\mcitedefaultseppunct}\relax
\EndOfBibitem
\bibitem[Simha and Ramaswamy(2002)]{simha2002hydrodynamic}
R.~A. Simha and S.~Ramaswamy, \emph{Physical Review Letters}, 2002,
  \textbf{89}, 058101\relax
\mciteBstWouldAddEndPuncttrue
\mciteSetBstMidEndSepPunct{\mcitedefaultmidpunct}
{\mcitedefaultendpunct}{\mcitedefaultseppunct}\relax
\EndOfBibitem
\bibitem[Bhattacharyya and Yeomans(2021)]{bhattacharyya2021coupling}
S.~Bhattacharyya and J.~M. Yeomans, \emph{Soft Matter}, 2021, \textbf{17},
  10716--10722\relax
\mciteBstWouldAddEndPuncttrue
\mciteSetBstMidEndSepPunct{\mcitedefaultmidpunct}
{\mcitedefaultendpunct}{\mcitedefaultseppunct}\relax
\EndOfBibitem
\bibitem[Giomi(2015)]{giomi2015geometry}
L.~Giomi, \emph{Physical Review X}, 2015, \textbf{5}, 031003\relax
\mciteBstWouldAddEndPuncttrue
\mciteSetBstMidEndSepPunct{\mcitedefaultmidpunct}
{\mcitedefaultendpunct}{\mcitedefaultseppunct}\relax
\EndOfBibitem
\bibitem[Nejad and Yeomans(2022)]{nejad2022active}
M.~R. Nejad and J.~M. Yeomans, \emph{Physical Review Letters}, 2022,
  \textbf{128}, 048001\relax
\mciteBstWouldAddEndPuncttrue
\mciteSetBstMidEndSepPunct{\mcitedefaultmidpunct}
{\mcitedefaultendpunct}{\mcitedefaultseppunct}\relax
\EndOfBibitem
\bibitem[Blow \emph{et~al.}(2014)Blow, Thampi, and Yeomans]{blow2014biphasic}
M.~L. Blow, S.~P. Thampi and J.~M. Yeomans, \emph{Physical Review Letters},
  2014, \textbf{113}, 248303\relax
\mciteBstWouldAddEndPuncttrue
\mciteSetBstMidEndSepPunct{\mcitedefaultmidpunct}
{\mcitedefaultendpunct}{\mcitedefaultseppunct}\relax
\EndOfBibitem
\bibitem[Dell’Arciprete \emph{et~al.}(2018)Dell’Arciprete, Blow, Brown,
  Farrell, Lintuvuori, McVey, Marenduzzo, and Poon]{dell2018growing}
D.~Dell’Arciprete, M.~Blow, A.~Brown, F.~Farrell, J.~S. Lintuvuori, A.~McVey,
  D.~Marenduzzo and W.~C. Poon, \emph{Nature Communications}, 2018, \textbf{9},
  1--9\relax
\mciteBstWouldAddEndPuncttrue
\mciteSetBstMidEndSepPunct{\mcitedefaultmidpunct}
{\mcitedefaultendpunct}{\mcitedefaultseppunct}\relax
\EndOfBibitem
\bibitem[Doostmohammadi \emph{et~al.}(2016)Doostmohammadi, Thampi, and
  Yeomans]{doostmohammadi2016defect}
A.~Doostmohammadi, S.~P. Thampi and J.~M. Yeomans, \emph{Physical Review
  Letters}, 2016, \textbf{117}, 048102\relax
\mciteBstWouldAddEndPuncttrue
\mciteSetBstMidEndSepPunct{\mcitedefaultmidpunct}
{\mcitedefaultendpunct}{\mcitedefaultseppunct}\relax
\EndOfBibitem
\bibitem[Streichan \emph{et~al.}(2018)Streichan, Lefebvre, Noll, Wieschaus, and
  Shraiman]{streichan2018global}
S.~J. Streichan, M.~F. Lefebvre, N.~Noll, E.~F. Wieschaus and B.~I. Shraiman,
  \emph{Elife}, 2018, \textbf{7}, e27454\relax
\mciteBstWouldAddEndPuncttrue
\mciteSetBstMidEndSepPunct{\mcitedefaultmidpunct}
{\mcitedefaultendpunct}{\mcitedefaultseppunct}\relax
\EndOfBibitem
\bibitem[He \emph{et~al.}(2014)He, Doubrovinski, Polyakov, and
  Wieschaus]{he2014apical}
B.~He, K.~Doubrovinski, O.~Polyakov and E.~Wieschaus, \emph{Nature}, 2014,
  \textbf{508}, 392--396\relax
\mciteBstWouldAddEndPuncttrue
\mciteSetBstMidEndSepPunct{\mcitedefaultmidpunct}
{\mcitedefaultendpunct}{\mcitedefaultseppunct}\relax
\EndOfBibitem
\bibitem[Mason \emph{et~al.}(2013)Mason, Tworoger, and Martin]{mason2013apical}
F.~M. Mason, M.~Tworoger and A.~C. Martin, \emph{Nature Cell Biology}, 2013,
  \textbf{15}, 926--936\relax
\mciteBstWouldAddEndPuncttrue
\mciteSetBstMidEndSepPunct{\mcitedefaultmidpunct}
{\mcitedefaultendpunct}{\mcitedefaultseppunct}\relax
\EndOfBibitem
\bibitem[Jagiella \emph{et~al.}(2016)Jagiella, M{\"u}ller, M{\"u}ller,
  Vignon-Clementel, and Drasdo]{jagiella2016inferring}
N.~Jagiella, B.~M{\"u}ller, M.~M{\"u}ller, I.~E. Vignon-Clementel and
  D.~Drasdo, \emph{PLoS Computational Biology}, 2016, \textbf{12},
  e1004412\relax
\mciteBstWouldAddEndPuncttrue
\mciteSetBstMidEndSepPunct{\mcitedefaultmidpunct}
{\mcitedefaultendpunct}{\mcitedefaultseppunct}\relax
\EndOfBibitem
\bibitem[Delarue \emph{et~al.}(2013)Delarue, Montel, Caen, Elgeti, Siaugue,
  Vignjevic, Prost, Joanny, and Cappello]{delarue2013mechanical}
M.~Delarue, F.~Montel, O.~Caen, J.~Elgeti, J.-M. Siaugue, D.~Vignjevic,
  J.~Prost, J.-F. Joanny and G.~Cappello, \emph{Physical Review Letters}, 2013,
  \textbf{110}, 138103\relax
\mciteBstWouldAddEndPuncttrue
\mciteSetBstMidEndSepPunct{\mcitedefaultmidpunct}
{\mcitedefaultendpunct}{\mcitedefaultseppunct}\relax
\EndOfBibitem
\bibitem[Delarue \emph{et~al.}(2014)Delarue, Joanny, J{\"u}licher, and
  Prost]{delarue2014stress}
M.~Delarue, J.-F. Joanny, F.~J{\"u}licher and J.~Prost, \emph{Interface Focus},
  2014, \textbf{4}, 20140033\relax
\mciteBstWouldAddEndPuncttrue
\mciteSetBstMidEndSepPunct{\mcitedefaultmidpunct}
{\mcitedefaultendpunct}{\mcitedefaultseppunct}\relax
\EndOfBibitem
\bibitem[Ranft \emph{et~al.}(2010)Ranft, Basan, Elgeti, Joanny, Prost, and
  J{\"u}licher]{ranft2010fluidization}
J.~Ranft, M.~Basan, J.~Elgeti, J.-F. Joanny, J.~Prost and F.~J{\"u}licher,
  \emph{Proceedings of the National Academy of Sciences}, 2010, \textbf{107},
  20863--20868\relax
\mciteBstWouldAddEndPuncttrue
\mciteSetBstMidEndSepPunct{\mcitedefaultmidpunct}
{\mcitedefaultendpunct}{\mcitedefaultseppunct}\relax
\EndOfBibitem
\end{mcitethebibliography}
\end{document}